\documentclass{article}
\usepackage{hiph-preprint}
\usepackage{graphicx}
\volnumber{22} \issuenumber{1} \edyear{2005}                             
\frompage{000} \topage{000}                                              
\recrevdate{18 November 2005}                                              
\def\rightmark{Gauge-field fluctuations in color superconductors}
\def\leftmark{J.L.\ Noronha {\it et. al.}}

\title{Effect of gauge-field fluctuations
on the phase transition between normal and color-superconducting
quark matter}
\authors{
{J.L.\ Noronha$^1$, H-c.\ Ren$^{2,3}$, I.\ Giannakis$^{2}$, D.\ Hou$^{3}$, and D.H.\ Rischke$^{1,4}$ %
\index{One, A.} 
\index{Two, A.} 
}\\[2.812mm]
{\normalsize
\hspace*{-8pt}$^1$ Frankfurt
Institute for Advanced Studies, J. W.
Goethe--Universit\"at, D-60438 Frankfurt am Main, Germany\\[0.2ex]
\hspace*{-8pt}$^2$ Physics
Department, The Rockefeller University,\\
1230 York Avenue, New
York, NY 10021-6399, USA\\
\hspace*{-8pt}$^3$ Institute of
Particle Physics, Huazhong Normal University,\\
430079 Wuhan, China\\
\hspace*{-8pt}$^4$ Institut
f\"ur Theoretische Physik, J. W. Goethe--Universit\"at,\\
D-60438 Frankfurt am Main, Germany\\
}}
 
\abstract{Type-I color superconductors display a first-order phase transition due to thermal gauge-field fluctuations. We numerically evaluated the critical temperature of the first-order phase transition and the corresponding discontinuity of the diquark condensate at the critical point.}


\makeindex
\begin{document}

\maketitle

\section{Introduction}\label{intro}
Quark matter at high baryon densities and sufficiently low temperatures is expected to be a color superconductor \cite{reviews}. The physics in
this region of the QCD phase diagram is relevant to explain the properties of highly compressed nuclear matter inside compact stars. Due to asymptotic freedom of QCD, color superconductivity can be quantitatively explored within QCD at asymptotic densities.

An analytical study concerning the effect of gauge fluctuations on the
free energy of a homogeneous type-I color-flavor locked (CFL) superconductor, near the critical temperature, was carried out in Ref.\ \cite{dirkfluct}. In this letter numerical results are presented for the fluctuation-induced critical temperature and the value of the diquark condensate at the transition temperature.

\section{Ginzburg-Landau free energy functional including gluon fluctuations}\label{techno}
The relevant Ginzburg-Landau (GL) free energy density including the gauge field fluctuations, $\gamma(t, \Delta)$, obtained in Ref.\ \cite{dirkfluct} as a function of the CFL gap function $\Delta(T)$ near $T_c$ reads
\begin{eqnarray}
\lefteqn{ \gamma(t, \Delta) \equiv  \frac{6\mu^2}{\pi^2}\,t \, \Delta^2(T)
+\frac{21\zeta(3)}{4\pi^4}\left(\frac{\mu}{k_BT_c}\right)^2\Delta^4(T)
}  \nonumber \\
& +&
32\pi(k_BT_c)^4\, F\left(\frac{1}{4\pi^2k_B^2T_c^2\lambda^2(T)}\right)\; ,
\label{free}
\end{eqnarray}
where $t\equiv (T-T_c)/T_c$ is the reduced temperature and $\mu$ is the quark chemical potential. The first two terms in eq. (\ref{free}) correspond to the GL free energy without the gluon fluctuations \cite{IidaIoannis}. The magnetic penetration depth $\lambda(T)$ is defined as $1 / \lambda^2(T)=\frac{7\zeta(3)}{24\pi^4} \left(
\frac{g\mu\Delta(T)}{k_BT_c}\right)^2$ and the last term
\begin{equation}
F(z)={\int^{\infty}_0}dx\, x^2 \left\{ \ln \left[1+{\frac{z}{x^2}}f(x)\right]
-\frac{z}{x^2}f(x) \right\}\;,
\label{eqrigas}
\end{equation}
reflects the contribution of the gluon fluctuations to the free energy. The derivation of this free energy will be presented in detail in a forthcoming publication \cite{future}. The function $f(y)$ describes the momentum dependence of the gluon magnetic mass in the static limit
\begin{equation}
f(y)=\frac{6}{7{\zeta}(3)}{\sum_{s=0}^{\infty}}{\int^{1}_0}dx
\frac{1-x^2}{(s+{\frac{1}{2}})[4(s+{\frac{1}{2}})^2+y^2x^2]}\;.
\label{eqkirkos}
\end{equation}
The limiting behavior of this function is $f(0)=1$ and
$f(y) \simeq 3{\pi}^3/[28{\zeta}(3)\,y]$ for $y \gg 1$. In terms of the coherence length at zero temperature ($\xi_0\equiv\frac{1}{2\pi k_BT_c}$) the variable $z(T)$ in eq. (\ref{eqrigas}) can be expressed as $z(T)=\frac{\xi^2_0}{\lambda^2(T)}$. This quantity determines for
which temperatures the local-coupling approximation of the
interactions between the gluons and the condensate is valid \cite{future}.

\section{Numerical results and comments}\label{maths}

The new critical temperature $T_c^*$ for the first-order phase transition and the value of the gap
$\Delta(T_c^*)$ were obtained from the nontrivial solution of the pair of equations
\begin{equation}
\gamma(t^*,\Delta) \equiv 0\quad, \qquad
\frac{\partial \gamma (t^*, \Delta)}{\partial \Delta^2} \equiv 0.
\label{eqwarp}
\end{equation}
We solved these equations numerically in order to find $\Delta(T_c^*)$ and $T_c^*$
as functions of $\mu$. In the Appendix it is explained how $T_{c}$ and $g$ were computed as functions of $\mu$. Fig.\ 1 (a) shows the ratio between $T_c^*$ and the pairing temperature $T_c$ at ultra-high chemical potentials.
\begin{figure}[htb]
\begin{center}
\includegraphics[width=0.58\textwidth,angle=270]{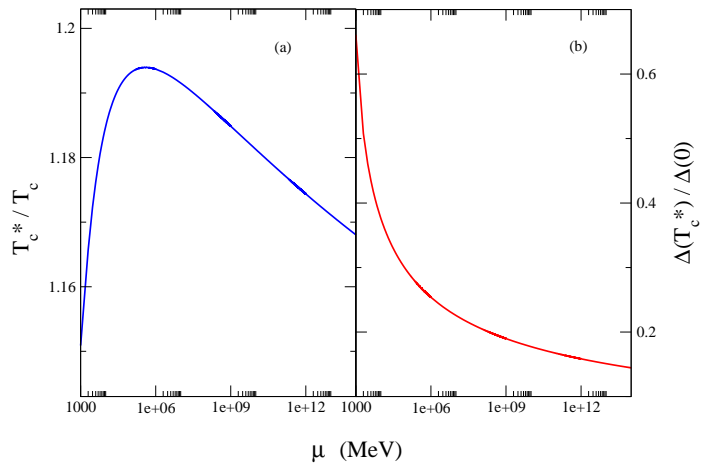}
\end{center}
\vspace*{-.2cm}
\caption[]{
(a) Comparison between the critical temperatures at high densities. (b) Discontinuity of the gap at the transition.
}
\label{fig1}
\end{figure}
The discontinuity of the gap at $T_c^*$ relative to its value at $T=0$ is shown in Fig.\ 1 (b). Note that $T_c^* > T_c$ at high, though finite, densities. Since $\frac{\Delta(T_c^*)}{\Delta(0)}\neq 0$ for nonzero chemical potentials we always have a first-order phase transition. However, both plots indicate that for $\mu \rightarrow \infty$ we have $\frac{T_c^*}{T_c} \rightarrow 1$ and $\frac{\Delta(T_c^*)}{\Delta(0)} \rightarrow 0$, meaning that a second-order phase transition takes place at infinitely large density. Our results for $T_c^*$ and $\Delta(T_c^*)$ are different than what was found in Ref.\ \cite{baym}. Matsuura et al. obtained that the fluctuation-induced critical temperature was smaller than $T_c$ and the ratio $\frac{\Delta(T_c^*)}{\Delta(0)}$ was found to be much smaller than what is shown in Fig.\ 1 (b). The reason why our results do not agree with those obtained in Ref.\ \cite{baym} is that in our approach the gluon fluctuations contribute always negatively to the GL free energy, which favors the diquark condensation and increases the corresponding fluctuation induced critical temperature \cite{future}.

In summary, we showed that gauge-field fluctuations not only induce a strong first-order phase transition but also increase the corresponding critical temperature beyond its mean-field value.
\section*{Appendix}\label{app}
In this paper we used the temperature $T_c$ obtained in Ref.\ \cite{son}, $\ln\frac{k_BT_c}{\mu}=-\frac{3\pi^2}{\sqrt{2}g}
+\ln\frac{2048\sqrt{2}\pi^3}{(N_{f}g^2)^\frac{5}{2}}
+\gamma-\frac{\pi^2+4}{8}$, which corresponds to the pairing instability of the normal phase. We also used the 3-loop formula for $\alpha=\frac{g^2}{4\pi}$ to compute $g(\mu)$ \cite{pdg}, $\alpha(\mu) = \frac{4\pi}{\beta_{0}\ln(\mu^2/\Lambda^2)}\Big[1-\frac{2\beta_1}{\beta_{0}^2}\frac{\ln(\ln(\mu^2/\Lambda^2))}{\ln(\mu^2/\Lambda^2)}  	+\frac{4\beta_{1}^2}{\beta_{0}^4(\ln(\mu^2/\Lambda^2))^2}\left( (\ln(\ln(\mu^2/\Lambda^2))-1/2)^2 +\frac{\beta_{2}\beta_{0}}{8\beta_{1}^{2}} -\frac{5}{4}    \right)\Big]$, where $\beta_0 = 9$, $\beta_1=51-\frac{19}{3}N_f=32$, $\beta_2=2857-\frac{5033N_f}{9}+\frac{325N_{f}^{2}}{27}$, $\Lambda=364$, for three quark flavors $N_f=3$ and three colors.
\section*{Acknowledgments}
J.L.N. thanks the Frankfurt International Graduate School for Science (FIGSS) for financial support. The work of I.G. and H.C.R. was supported in part by the US Department of Energy under grants DE-FG02-91ER40651-TASKB while the work of D.H. was partly supported by the NSFC under grant No. 10135030 and the Educational Committee of China under grant No. 704035. H.C.R. and D.H. were also partly supported by the NFSC under grant No. 10575043.

\vfill\eject
\end{document}